\documentstyle[preprint,aps]{revtex}

\begin{document}
\title{Decoherence at zero temperature}
\author{G. W. Ford$\dagger$ and R. F. O'Connell$\ddagger$}
\address{School of Theoretical Physics, Dublin Institute for\\
Advanced Studies, \\
10 Burlington Road, Dublin 4, Ireland}
\date{\today}
\maketitle

\begin{abstract}
Most discussions of decoherence in the literature consider the
high-temperature regime but it is also known that, in the presence of
dissipation, decoherence can occur even at zero temperature. Whereas
most
previous investigations all assumed initial decoupling of the quantum
system
and bath, we consider that the system and environment are entangled at
all
times. Here, we discuss decoherence for a free particle in an initial
Schr\"{o}dinger cat state. Memory effects are incorporated by use of the
single
relaxation time model (since the oft-used Ohmic model does not give
physically correct results).
\end{abstract}


Most discussions of decoherence obtain results for decoherence decay
times
which are proportional to the inverse of $T$ or $T^{1/2}$, where $T$ is
the
temperature, and which apply only to the high temperature regime,
$kT>>\hbar
\gamma $, where $\gamma $ is a typical dissipative decay rate \cite
{giulini,walls,zurek,ford,ford2}. In fact, whereas most papers in the
early
literature give decay rates proportional to $\gamma ^{-1}$, we showed
that
decoherence can occur at high temperature even for vanishingly small
dissipation\cite{ford,ford2}. For a pedagogical discussion of the high
$T$
regime, we refer to \cite{ford02}. Here, we look at the opposite end of
the
temperature spectrum and present explicit results at temperature zero.
In
contrast to the high temperature regime, we find that a dissipative
environment is necessary to achieve decoherence.

The zero temperature case was also considered by others, most notably by
Walls and Milburn\cite{walls}. They considered a "Schr\"{o}dinger cat"
state
consisting of a pair of coherent states of an oscillator. They solved
the
master equation to obtain an expression for the coherence attenuation
coefficient, defined in terms of the off-diagonal elements of the
density
matrix. However, this approach is subject to serious assumptions: (a)
initial decoupling of the system and environment; (b) weak coupling,
which
implies a restriction to a finite oscillator potential.

The assumption (a) used by Walls and Milburn and indeed many other
investigators, leads to results which are seriously at variance with a
model
in which the quantum particle and the heat bath are entangled at all
times..
In fact, at $T=0$, we are simply dealing with the zero-point
oscillations of
the heat bath which are necessarily {\em always entangled} with whatever
quantum particle is of interest. In fact, the results we obtain here are
applicable to the whole low temperature regime, defined by the
inequality $
kT<<\hbar \gamma $. General results for all regimes, corresponding to
arbitrary $T$ and $\gamma $, were obtained in our initial paper on this
subject \cite{ford}, where our starting-point was the {\em density
matrix}
for the {\em system} of particle coupled to the bath at all times.
However,
there explicit results were presented only for the Ohmic model in the
high
temperature regime ($kT>>\hbar \gamma $). Later, we showed that the same
results could be obtained simply by starting with the wave function
describing a pure Schr\"{o}dinger cat state moving with an initial
velocity $v$\cite{ford2}. This was then used to calculate the wave
function and then the coordinate probability distribution at time $t$.
Finally, this probability
distribution was then averaged over a thermal distribution of initial
velocities to obtain the probability distribution corresponding to a
finite
high temperature $T$. This led immediately to a quantitative measure of
decoherence (the destruction of the interference pattern). In other
words,
the calculation of the decoherence decay time at high temperature could
be
carried out simply within the framework of elementary quantum mechanics
and
equilibrium statistical mechanics, with a result independent of $\gamma
$
\cite{ford2,ford02} .

In contrast to the high temperature case, results at low temperature
depend
essentially on $\gamma $ and must be obtained within the framework of
nonequilibrium statistical mechanics. Thus, we will now give a synopsis
of
our general model describing a system and its environment as being
entangled
at all times \cite{ford}.

First, we emphasize that, not only are our results in general applicable
to
arbitrary dissipation and arbitrary temperature, but they are not
subject to
the weak coupling approximation and so apply in the case of a free
particle.
Arbitrary dissipation refers to reservoir models which give rise to
memory
terms in the quantum Langevin equation (see (\ref{4}) below and the
subsequent discussion) which describes the equation of the quantum
system in
a dissipative environment. By contrast, the Ohmic model (the model
considered by most authors) is the simplest dissipative model and
corresponds to the lack of memory terms in the Langevin equation. As we
shall see below, the Ohmic model has the difficulty that the mean square
velocity at zero temperature is divergent and a cutoff must be supplied
to
make it finite. For this purpose we use the single relaxation time model
(which reduces to the Ohmic model in the limit of zero relaxation time).

As in our previous discussions\cite{ford,ford2,ford02}, we consider
decoherence in terms of the simple problem of a free particle moving in
one
dimension that is placed in an initial superposition state
(\textquotedblleft Schr\"{o}dinger cat\textquotedblright\ state)
corresponding to a pair of Gaussian wave packets, each with variance
$\sigma
^{2}$ and separated by a distance $d\gg \sigma $. For such a state the
probability distribution at time $t$ can be shown to be of the form:
\begin{eqnarray}
P(x,t) &=&\frac{1}{2(1+e^{-d^{2}/8\sigma ^{2}})}\left\{
P_{0}(x-\frac{d}{2}
,t)+P_{0}(x+\frac{d}{2},t)\right.   \nonumber \\
&&\left. +2e^{-d^{2}/8w^{2}(t)}a(t)P_{0}(x,t)\cos \frac{[x(0),x(t)]xd}{
4i\sigma ^{2}w^{2}(t)}\right\} ,  \label{0}
\end{eqnarray}
where $P_{0}$ is the probability distribution for a single wave packet,
given by
\begin{equation}
P_{0}(x,t)=\frac{1}{\sqrt{2\pi w^{2}(t)}}\exp
\{-\frac{x^{2}}{2w^{2}(t)}\}.
\label{0a}
\end{equation}
Here and in (\ref{0}) $w^{2}(t)$ is the variance of a single wave packet,
which in general is given by
\begin{equation}
w^{2}(t)=\sigma ^{2}-{\frac{[x(0),x(t)]^{2}}{4\sigma ^{2}}}+s(t),
\label{1}
\end{equation}
where $\sigma ^{2}$ is the initial variance, $[x(0),x(t)]$ is the
commutator, and
\begin{equation}
s(t)=\langle \{x(t)-x(0)\}^{2}\rangle ,  \label{2}
\end{equation}
is the mean square displacement. In (\ref{0}) the first two terms within
the
braces correspond to the two wave packets, centered at $\pm d/2$,
expanding
independently, while the third term is the interference term.
Decoherence
refers to the destruction of interference, a measure of which is given
by
the attenuation coefficient $a(t)$ which can be defined as the ratio of
the
factor multiplying the cosine in the interference term to twice the
geometric mean of the first two terms \cite{ford,ford2}. Using a method
of
successive measurements introduced by Ford and Lewis \cite{ford86}, in
which
a particle in equilibrium and entangled with the environment is placed
in
the initial state by a first measurement and then, after a time interval
$t$
, is probed by a second measurement, we obtained the following exact
general
formula for the attenuation coefficient \cite{ford,ford86}.
\begin{equation}
a(t)=\exp \{-{\frac{s(t)d^{2}}{8\sigma ^{2}w^{2}(t)}}\}.  \label{3}
\end{equation}

The quantities appearing in (\ref{1}) and (\ref{2}) are evaluated by use
of
the quantum Langevin equation \cite{ford88}, which is a Heisenberg
equation
of motion for $x(t)$, the dynamical variable corresponding to the
coordinate
of a Brownian particle interacting with a linear passive heat bath. For
the
case of a free particle, this equation for the stationary process has
the
well known form,
\begin{equation}
m\ddot{x}+\int_{-\infty }^{t}dt^{\prime }\mu (t-t^{\prime })\dot{x}
(t^{\prime })=F(t),  \label{4}
\end{equation}
where $\mu (t)$ is the memory function and $F(t)$ is a fluctuating
operator
force with mean zero. The solution of the quantum Langevin equation
(\ref{4}
) can be written
\begin{equation}
x(t)=\int_{-\infty }^{t}dt^{\prime }G(t-t^{\prime })F(t^{\prime }),
\label{5}
\end{equation}
where $G(t)$, the Green function can in turn be written
\begin{equation}
G(t)=\frac{1}{2\pi }\int_{-\infty }^{\infty }d\omega \alpha (\omega
+i0^{+})e^{-i\omega t},  \label{6}
\end{equation}
in which $\alpha (z)$ (the Fourier transform of the Green function) is
the
response function. For the free particle the response function has
the
general form
\begin{equation}
\alpha (z)=\frac{1}{-mz^{2}-iz\tilde{\mu}(z)},  \label{6b}
\end{equation}
in which $\tilde{\mu}(z)$ is the Fourier transform of the memory
function,
\begin{equation}
\tilde{\mu}(z)=\int_{0}^{\infty }dt\mu (t)e^{izt},\qquad {\rm
Im}\{z\}>0.
\label{6c}
\end{equation}

Using these results, we find that \cite{ford88,ford89} the mean square
displacement is given by the formula
\begin{equation}
s(t)=\frac{2\hbar }{\pi }\int_{0}^{\infty }d\omega {\rm Im}\{\alpha
(\omega
+i0^{+})\}\coth \frac{\hbar \omega }{2kT}(1-\cos \omega t),  \label{7}
\end{equation}
while the commutator, which is temperature independent, is given by the
formula
\begin{equation}
\lbrack x(t_{1}),x(t_{1}+t)]=\frac{2i\hbar }{\pi }\int_{0}^{\infty
}d\omega
{\rm Im}\{\alpha (\omega +i0^{+})\}\sin \omega t.  \label{8}
\end{equation}
These expressions are valid for arbitrary temperature and arbitrary
dissipation. (Indeed, with the appropriate expression for the response
function, they are valid in the presence of an external oscillator
potential.) In our earlier discussions, for brevity, we confined our
attention to the case of high temperature. Here we consider the case of
zero
temperature. Most discussions in the literature are confined to the
so-called Ohmic case, where $\tilde{\mu}(z)=\zeta $, the Newtonian
friction
constant but we analyze the more general case of the single relaxation
time
model \cite{ford01}. This model corresponds to a memory function of the
form
\begin{equation}
\mu (t)=\frac{\zeta }{\tau }e^{-t/\tau }\theta (t),  \label{10}
\end{equation}
where $\tau $ is the relaxation time of the bath and where $\theta $ is
the
Heaviside function. Note that in the limit $\tau \rightarrow 0$ this
becomes
the Ohmic memory function $\mu (t)=2\zeta \delta (t)\theta (t)$. With
this
form of the memory function,
\begin{equation}
\tilde{\mu}(z)=\frac{\zeta }{1-iz\tau }.  \label{11}
\end{equation}
Next, with this form in the expression (\ref{6b}) for the response
function,
we find with $T=0$ that the expression (\ref{7}) for the response can be
expressed in the form,
\begin{equation}
s(t)=\frac{2\hbar }{\pi \zeta }\frac{\Omega ^{2}V(\gamma t)-\gamma
^{2}V(\Omega t)}{\Omega ^{2}-\gamma ^{2}},  \label{12}
\end{equation}
while the expression (\ref{8}) for the commutator becomes
\begin{equation}
\lbrack x(0),x(t)]=\frac{i\hbar }{\zeta }\frac{\Omega ^{2}(1-e^{-\gamma
t})-\gamma ^{2}(1-e^{-\Omega t})}{\Omega ^{2}-\gamma ^{2}},  \label{13}
\end{equation}
where we have introduced the quantities,
\begin{equation}
\Omega =\frac{1+\sqrt{1-4\zeta \tau /m}}{2\tau },~~\gamma
=\frac{1-\sqrt{
1-4\zeta \tau /m}}{2\tau }.  \label{14}
\end{equation}
and where \cite{bateman}
\begin{eqnarray}
V(x) &=&\int_{0}^{\infty }dy\frac{x^{2}}{y(y^{2}+x^{2})}(1-\cos y)
\nonumber
\\
&=&\log x+\gamma _{E}-\frac{1}{2}[e^{-x}{\rm \bar{E}i}(x)+e^{x}{\rm
Ei}(-x)]
\nonumber \\
&=&-(\log x+\gamma _{E})(\cosh x-1)-\frac{1}{2}[e^{-x}\sum_{n=1}^{\infty
}
\frac{x^{n}}{n!n}+e^{x}\sum_{n=1}^{\infty }\frac{(-x)^{n}}{n!n}].
\label{15}
\end{eqnarray}
Here $\gamma _{E}=0.577215665$ is Euler's constant. Note that the sums
in (
\ref{15}) are absolutely convergent. We should point out that the
quantities
$\Omega $ and $\gamma $ were referred to as $\gamma _{+}$ and $\gamma
_{-}$
in \cite{ford01}. We feel that the present choice is more suggestive
since,
for $\tau \rightarrow 0$, we have $\Omega \rightarrow \frac{1}{\tau }$
and $
\gamma \rightarrow \zeta /m$.

Note the expansions \cite{batemanhtf}, for small $x$,
\begin{equation}
V(x)\cong -\frac{1}{2}x^{2}\left( \log x+\gamma _{E}-\frac{3}{2}\right)
.
\label{16}
\end{equation}
Also, asymptotically, for large $x,$
\begin{equation}
V(x)\sim \log x+\gamma
_{E}-\frac{1}{x^{2}}-\frac{3!}{x^{4}}-\frac{5!}{x^{6}}
.  \label{17}
\end{equation}
For very short times $(t<<\tau )$, one can show in general that
\begin{equation}
s(t)\cong \langle v^{2}\rangle t^{2},  \label{18}
\end{equation}
where $\langle v^{2}\rangle $ is the mean square velocity, given for the
single relaxation time model by
\begin{equation}
\langle v^{2}\rangle =\frac{\hbar \gamma \Omega }{\pi m(\Omega -\gamma
)}
\log \frac{\Omega }{\gamma }\cong -\frac{\hbar \zeta }{\pi m^{2}}\log
\frac{
\zeta \tau }{m}.  \label{19}
\end{equation}
We note that, for $\tau \rightarrow 0$ (corresponding to the Ohmic
model),
the mean square velocity, given in (\ref{19}), is logarithmically
divergent.
The conclusion is that, in order to obtain finite results at short time,
one
must use a cutoff model, such as the single relaxation time model used
here.

For intermediate times ($\tau <<t<<(\zeta /m)^{-1}$), we obtain
\begin{equation}
s(t)\cong -\frac{\hbar \zeta }{\pi m^{2}}t^{2}\left\{ \log \frac{\zeta
t}{m}
+\gamma _{E}-\frac{3}{2}\right\} .  \label{20}
\end{equation}
We can compare these results with the weak coupling result which for a
free
particle leads to $s(t)$ being identically zero. (For a free particle
the
mean square velocity vanishes at zero temperature.)

Thus, to investigate the decay of coherence (which is generally a
short-time
phenomena) in both time ranges, we consider the behavior for times
$\zeta
t/m<<1$  In this case, the commutator becomes that of a non-interacting
particle,
\begin{equation}
\lbrack x(0),x(t)]\cong i\frac{{\hbar }t}{m}.  \label{21}
\end{equation}
Hence, it is clear that for very short times the wave packet width is
the
initial width,
\begin{equation}
w^{2}(t)\cong \sigma ^{2}.  \label{22}
\end{equation}
Thus, the attenuation coefficient (\ref{3}) for short times takes the
form:
\begin{equation}
a(t)\cong \exp \left\{ -\frac{s(t)d^{2}}{8\sigma ^{4}}\right\} ,
\label{23}
\end{equation}
where $s(t)$ is given by (\ref{18}) and (\ref{20}). Hence, in
particular, we
can write
\begin{equation}
a(t)=\exp \left\{ \left( \frac{t}{\tau _{0}}\right) ^{2}\log \frac{\zeta
\tau }{m}\right\} ,~~~~~~t<<\tau   \label{24}
\end{equation}
and

\begin{equation}
a(t)=\exp\left\{\left(\frac{t}{\tau_{0}}\right)^{2}\left[\log\frac{\zeta
t}{m}+\gamma_{E}-\frac{3}{2}\right]]\right\},~~\tau <<t<<(\zeta/m)^{-1}
\label{28}
\end{equation}
with

\begin{equation}
\tau _{0}\equiv \frac{m\sigma ^{2}}{d}\sqrt{\frac{8\pi }{\hbar \zeta }}.
\label{25}
\end{equation}
The characteristic time for decoherence to occur, $\tau _{d}$, is
defined as
usual \cite{ford,ford2,ford02}, as the time at which $a(t)=\exp
(-1)$
. Hence, for example it follows from (\ref{24}) that, for $t<<\tau$,
\begin{equation}
\tau _{d}=\tau _{0}|\log (\zeta \tau /m)|^{-1/2}~~~~~~<\tau _{0},
\label{26}
\end{equation}
where the inequality sign follows from the fact that $\zeta t/m<<1$.  It
is also clear that $\tau_{d}<\tau_{0}$ in the case $t>>\tau$.

Next, we consider some numerical results. As mentioned above, here
$\zeta
/m\approx \gamma $ and, as a point of reference, it is useful to note
that
\begin{equation}
\frac{kT}{\hbar \gamma }=\frac{T(K)}{\gamma (10^{11}s^{-1})}.
\label{27}
\end{equation}
Now, one of the few experimental investigations of decoherence is the
work
of Myatt et al. on the decoherence of trapped $^{9}Be^{+}$ ions
\cite{myatt}
for which $\gamma $ values $\approx 6\times 10^{3}$ are quoted. It is
clear
from (\ref{27}) that for such low values of $\gamma $ that we are in the
low-temperature regime even for $T$ values which would normally be
considered very high. Finally, as an illustration, if we consider a
$^{9}Be$
ion  with $\gamma\approx 6\times 10^{3}$ and if we take $\sigma
=1\AA$ and
$d=1cm$ then, from (\ref{25})
we obtain a $\tau _{0}$ value of $\approx 6\times 10^{-16}s^{-1}$ and,
as
mentioned above, $\tau _{d}$ will be even smaller i.e. decoherence
occurs in
a very short time even at zero temperature and indeed in the whole low
temperature regime.

\acknowledgments{The authors wish
to thank the School of Theoretical Physics, Dublin Institute
for Advanced Studies, for their hospitality.}

\end{document}